%% file: floquet_dressing_resub.tex
\documentclass[aps,pra,groupedaddress,showpacs,letterpaper,twocolumn,reprint,nofootinbib,longbibliography]{revtex4-1}

\usepackage{graphicx}
\usepackage{setspace}
\usepackage{gensymb}
\usepackage{braket}
\usepackage{amsmath}
\usepackage{color}

\begin{document}

\newcommand{\GHzpVpcm}[0]{\mathrm{GHz\;(V/cm)^{-2}}}
\newcommand{\mVpcm}[0]{mV/cm}
\newcommand{\Vpcm}[0]{V/cm}
\newcommand{\Vpcmsq}[0]{V\;cm^{-2}}
\newcommand{\numunit}[2]{$#1 \;\mathrm{#2}$}
\newcommand{\numunitexp}[3]{$#1 \times 10^{#2} \;\mathrm{#3}$}
\newcommand{\numunitunc}[3]{$#1 \pm #2 \;\mathrm{#3}$}
\newcommand{\state}[2]{#1_{#2}}

\newcommand {\editms}[1]{\textcolor{blue}{$\spadesuit$~#1}}

\title{Reducing the sensitivity of Rydberg atoms to dc electric fields using two-frequency ac field dressing}
\author{Donald W. Booth}
\author{Joshua Isaacs}
\author{M. Saffman}
\affiliation{Department of Physics, University of Wisconsin-Madison, 1150 University Avenue, Madison, Wisconsin 53706}

 \date{\today}

\begin{abstract}
We propose a method for reducing the sensitivity of atomic ground to Rydberg transitions  to stray dc electric fields, using microwave-induced dressing of Rydberg states. Calculations are presented for  the Cs $\state{90S}{1/2}$ and $\state{90P}{3/2}$ states. With zero dc bias electric field, a two-frequency ac field is used to simultaneously reduce the sensitivity of both states to dc field variations. The sensitivity reduction is a factor of 95 for the $\state{90S}{1/2}$ state and a factor of 1600 for the  $\state{90P}{3/2}, m_J=3/2$ state. 
 We also show how the two-frequency ac field can be used to cancel  both second- and fourth-order terms in the polarizability of a single Rydberg state.
These results are relevant to improving the stability of  experiments that seek to excite Rydberg atoms in the proximity of charged surfaces. 
\end{abstract}

\maketitle

\section{Introduction}

One of the most promising approaches to quantum computation uses the strong interactions between Rydberg states of neutral atoms for quantum logic gates\cite{Saffman2010}. The two-atom interaction strength scales as $n^4/R^3$ for resonant interactions and $n^{11}/R^6 - n^{12}/R^6$ for long range van der Waals interactions\cite{Saffman2010} where $n$ is the principal quantum number and $R$ is the atomic spacing. Protocols for high fidelity quantum gates  are optimized for $n\sim 100$\cite{Saffman2016}  which  implies a large contrast between the ground-ground and Rydberg-Rydberg interaction strength.  The strong Rydberg-Rydberg interaction is due to the fact that the wave function of an electron in a  Rydberg state has a size that scales as $n^2 a_0$ with $a_0$ the Bohr radius. Unfortunately the large wavefunction  implies that the sensitivity of the atomic energy to electric fields, which is quantified by the polarizability
$\alpha$, also grows rapidly scaling as $\alpha\sim n^7$\cite{Gallagher1994}. 
 
Due to the large Rydberg state polarizability it is necessary to reduce any possible perturbations from background electric fields for stable operation of Rydberg state mediated gates. Such perturbations would otherwise result in detuning of the excitation laser from the ground to Rydberg transition giving gate errors. This is true  for the usual atom-atom gates and for protocols that aim to  create entanglement between atoms and microwave photons for hybrid interfaces\cite{Sorensen2004,Petrosyan2009,Hafezi2012,Pritchard2014,Sarkany2015} or between 
photons\cite{Firstenberg2016}.  The atom-photon gate, as described in \cite{Pritchard2014}, requires stable excitation from the ground to a Rydberg state and also requires that the frequency separation between neighboring Rydberg states is unperturbed to maintain resonance  with a microwave photon. Thus we require that the absolute polarizability of two Rydberg states is minimized.  Perturbations due to background fields are particularly problematic when placing atoms near surfaces, which is desirable in order to miniaturize the experimental platform\cite{Tauschinsky2010,Carter2013,Bernon2013} or to enhance coupling to microwave fields carried by planar waveguides\cite{Hogan2012,Thiele2014,Beck2016,Gard2017}.

In experiments with cold atoms or ions surface fields appear due to contamination by adsorbates. 
Several groups have measured and characterized the fields near surfaces using methods such as the motion of atoms in a Bose-Einstein condensate \cite{Obrecht2007}, heating of trapped ions\cite{Brownnutt2015}, Rydberg electromagnetically induced transparency \cite{Abel2011,Chan2014}, and Rydberg Stark spectroscopy \cite{Naber2016,Thiele2015}. These fields can have substantial gradients; for example in \cite{Hermann-Avigliano2014}, an electric field gradient of 12 V/cm$^2$ was observed above a superconducting atom chip. Furthermore, several attempts have been made to reduce the effects of these fields. These approaches include reducing the fields by using adsorbates to cancel stray fields \cite{Sedlacek2016} and baking the substrate to diffuse the adsorbates across the surface \cite{Obrecht2007}.

One potential approach to reducing the sensitivity of Rydberg atoms to stray dc, or slowly varying, electric fields uses microwave fields to admix atomic states to reduce the atoms' polarizabilities \cite{Hyafil2004,Mozley2005,Jones2013,Ni2015}. In \cite{Jones2013}, microwave fields at \numunit{\sim\nobreak 38}{GHz} are used to cancel the relative polarizabilities between the $\state{48S}{1/2}$ and $\state{49S}{1/2}$ Rydberg levels in $^{87}\mathrm{Rb}$ by coupling the $S$ states to neighboring $P$ states. The $P$ states have polarizabilities of the same sign as $S$ states and thus cannot cancel the  $S$ state polarizabilities, so the experiment aims only to cancel the relative Stark shift between two Rydberg levels. In \cite{Ni2015} this relative polarizability cancellation was extended to pairs of circular Rydberg states. It is also possible to control  other properties of Rydberg atoms using microwaves, such as the interaction strength between neighboring atoms \cite{Bohlouli-Zanjani2007,Sevincli2014,Tretyakov2014,XFShi2017}.

Previous work on reducing the Rydberg sensitivity with microwave dressing has only considered the problem of the differential polarizability of nearby Rydberg states\cite{Hyafil2004,Mozley2005,Jones2013,Ni2015}. Experiments that rely on resonant excitation of Rydberg states are also sensitive to the magnitude of the polarizability of a single Rydberg state. In this paper we show that  by dressing with two microwave frequencies we can greatly reduce the magnitude of the polarizability of neighboring opposite parity Rydberg states. In this way  ground-Rydberg  and  Rydberg-Rydberg transitions are simultaneously rendered insensitive to low-frequency electric field noise. This is particularly important for proposals that rely on stable ground-Rydberg and Rydberg-Rydberg transition energies, as in \cite{Pritchard2014}. In addition we show that if the goal is only to reduce the polarizability of a single state then two-frequencies can be used to cancel higher order terms of the hyperpolarizability. 

The remainder of the paper is organized as follows. In Sec. \ref{sec.method} we present the detailed design of the two-frequency dressing method. We use as an example the specific case of excitation of a Cs atom to the $90P_{3/2}$ state, and its coupling to $90S_{1/2}$. These choices are motivated by our proposal for  atom-microwave photon entanglement using these states\cite{Pritchard2014}.  In Sec. \ref{sec.results} we present numerical results from a Floquet analysis showing reduction of the polarizability and the dependence on the direction of the background field. In Sec. \ref{sec.summary} we summarize the results obtained.

\section{Two-frequency microwave dressing}
\label{sec.method}

\begin{figure}[!t]
    \centering
    \includegraphics[width=0.48\textwidth]{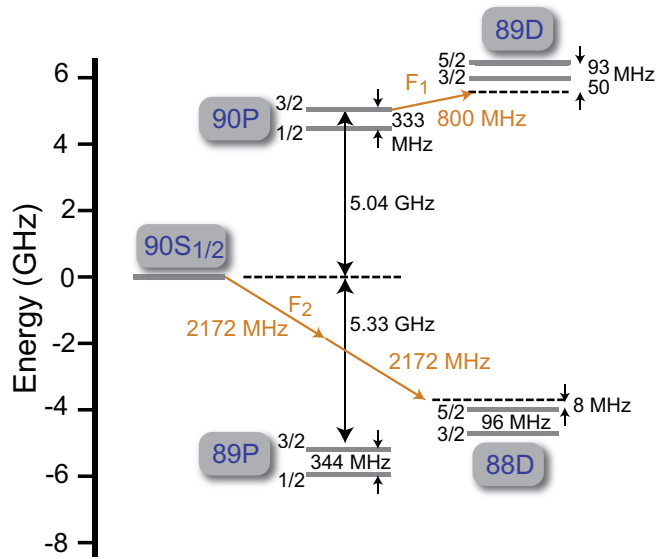}
    \caption{Level diagram and dressing fields for two-frequency polarizability nulling. The energy scale is relative to the $\state{90S}{1/2}$ state. Microwave fields with frequencies \numunit{800}{MHz} and \numunit{2172}{MHz}, depicted in the diagram as dotted arrows, couple the $\state{90P}{3/2}$ state off-resonantly to the $\state{89D}{5/2}$ state, and couples the $\state{90S}{1/2}$ state to the $\state{88D}{5/2}$ state by a nearly-resonant two-photon transition. }
    \label{fig:leveldiagram}
\end{figure}

We propose a method for cancelling the absolute polarizabilities of $nS$ and $nP$ Rydberg states by using linearly-polarized microwave fields to admix components of neighboring Rydberg states. We focus on the $\state{90S}{1/2}$ and $\state{90P}{3/2}$ states in Cs, and consider the effects on these states of a two-frequency microwave field as shown in Fig. \ref{fig:leveldiagram}. The scalar and tensor dc polarizabilities\cite{Khadjavi1968} of these states are 
$\alpha_0^{(90S_{1/2})}=3.501 ,~ \alpha_2^{(90S_{1/2})}=0, ~\alpha_0^{(90P_{3/2})}= 95.448,~ \alpha_2^{(90P_{3/2})}=-8.274 $ in units of $\rm GHz/(V/cm)^2$. We will only consider linearly polarized fields so the vector polarizabilities do not play a role. 
These values were calculated using a standard sum over states method\cite{Iskrenova-Tchoukova2007} with matrix elements calculated with a WKB approximation\cite{Kaulakys1995}  using quantum defects from \cite{Lorenzen1984,Weber1987}. Since the $90S_{1/2}$ and $90P_{3/2}$ states both have positive scalar polarizability, and an order of magnitude smaller tensor polarizability for the $90P_{3/2}$ state  the only way to null the polarizability is to admix $nD$ states that have negative scalar polarizability.

Admixing is achieved using a 800 MHz field $F_1$ to couple $90P_{3/2}$ to $89D_{5/2}$ with a one-photon transition and a 2172 MHz field $F_2$ to couple $90S_{1/2}$ to $88D_{5/2}$ with a two-photon transition. The selected frequencies result in near resonant couplings and were chosen to minimize the dc field sensitivity of $90S_{1/2}$ and $90P_{3/2}$ as explained below.   Because the microwave frequencies are far off-resonance from the 5.04 GHz $\state{90S}{1/2} \leftrightarrow \state{90P}{3/2}$ transition, any mixing between these states is small compared to the mixing with the $D$ states.

If the fields $F_1, F_2$ are sufficiently weak a perturbative calculation is sufficient to calculate the polarizability of the mixed states in the presence of the dressing fields. 
In \cite{Jones2013}, a perturbative approach was compared to a more accurate Floquet calculation. The validity of the perturbative method depends on the ratio of the Rabi frequency of the microwave field to the carrier frequency of the microwave field. The perturbative method is valid in cases where this ratio is $\ll 1$. In  \cite{Jones2013} this ratio was $\sim 1/30$, due to the relatively small transition dipoles at lower principal quantum number $n$ and the higher microwave frequencies used. In our case, the Rabi frequency is \numunit{\sim 3.5}{GHz}, which is of the same order as the frequency of the microwave fields, \numunit{800}{MHz} and \numunit{2172}{MHz}. Thus we must use the Floquet method rather than a perturbative approach.

To calculate the effects of the microwave fields on  Rydberg states, we implemented the Floquet method as described in \cite{Ni2015}.  We split the Hamiltonian into time-independent and time-dependent parts
\begin{eqnarray}
H &=& H_0 + H_{\rm ac}(t) , \nonumber\\
H_0 &=& H_{\rm a} + H_{\rm dc} , \nonumber \\
H_{\rm dc} &=& F_{\rm dc} z,\nonumber\\
H_{\rm ac}(t) &=& \sum_{i=1,2}F_{i} \cos\left(\omega_i t+\phi_i\right)z .\nonumber
\end{eqnarray}
The time-independent part $H_0$ contains two terms. The first term  $H_{\rm a}$ accounts for the  the field-free state energies
\begin{equation}
\bra{nlj}H_{\rm a}\ket{nlj}=E_{nlj} = - \frac{E_{\rm H}}{2(n-\delta_{nlj})^2}\nonumber
\end{equation}
where $E_{\rm H}$ is the Hartree energy and  $\delta_{nlj}$ are the quantum defects.

The second term  $H_{\rm dc}$ contains the dependence on the dc electric field. The dc Stark matrix is calculated using the method from \cite{Zimmerman1979}. The matrix elements of $H_{\rm dc}=F_{\rm dc} z $ are of the form (Eq. (10) in \cite{Zimmerman1979})
\begin{eqnarray}
 \braket{\alpha \left| F_{\rm dc} z \right| \beta} &=& \delta_{m_j,m_j'} \delta_{l,l' \pm 1} \braket{\alpha \left| r \right| \beta} F_{\rm dc} \nonumber\\
 \times \hspace{-.6cm}\sum_{m_l=m_j \pm 1/2} && \hspace{-.9cm}
C_{l,1/2,m_l,m_j-m_l}^{j,m_j}
C_{l',1/2,m_l,m_j-m_l}^{j',m_j} \braket{l, m_l \left| \mathrm{cos} \theta \right| l',m_l},\nonumber\\ 
\label{eq.Hdc}
\end{eqnarray}
where $\alpha$ and $\beta$ are shorthand for the quantum numbers  $\{n,l,j,m_j\}$, $\{n',l',j',m_j'\}$, $F_{\rm dc}$ is the dc electric field strength, and the first two factors in the sum are Clebsch-Gordan coefficients. The final factor in the sum evaluates to
\begin{align}
& \braket{l, m \left| \mathrm{cos} \theta \right| l-1,m} = \left[\frac{l^2-m^2}{(2l+1)(2l-1)}\right]^{1/2} , \nonumber \\
& \braket{l, m \left| \mathrm{cos} \theta \right| l+1,m} = \left[\frac{(l+1)^2-m^2}{(2l+3)(2l+1)}\right]^{1/2} \nonumber. 
\end{align}
Equation (\ref{eq.Hdc}) gives the matrix elements for a field aligned along the quantization axis. In Sec. \ref{sec.results} below we will also consider 
the effect of rotated fields. The matrix elements are then calculated using (\ref{eq.Hdc}) with Wigner $D$ functions to rotate the states. All calculations are performed without accounting for the hyperfine structure of the Rydberg states. The hyperfine splitting of Cs $90s$ is approximately 100 kHz and substantially smaller for $np, nd$ states. Since this is several orders of magnitude smaller than the detuning of the dressing fields the hyperfine structure gives only a very minor correction to the results. 

The time-dependent interactions due to the ac fields $F_1,F_2$ with frequencies and phases $\omega_{1,2}, \phi_{1,2}$ are contained 
in $H_{\rm ac}(t)$. In all calculations  the  ac field will be  polarized along
the $z$ direction.
Since these interactions are periodic in time, we can apply the Floquet method to this problem. The Floquet method is described in detail in \cite{Shirley1965}. The solutions of this system are represented in the form
\begin{equation}
{\bf F}(t) = \mathbf{\Phi}(t) e^{-i\mathbf{Q}t} ,\nonumber
\end{equation}
where $\mathbf{\Phi}(t)$ is a matrix of periodic functions and $\mathbf{Q}$ is a time-independent diagonal matrix, whose elements $q_\alpha$ are called the ``quasi-energies" for the modes of the system. 

\begin{figure*}[!ht]
    \centering
    \includegraphics[width=0.94\textwidth]{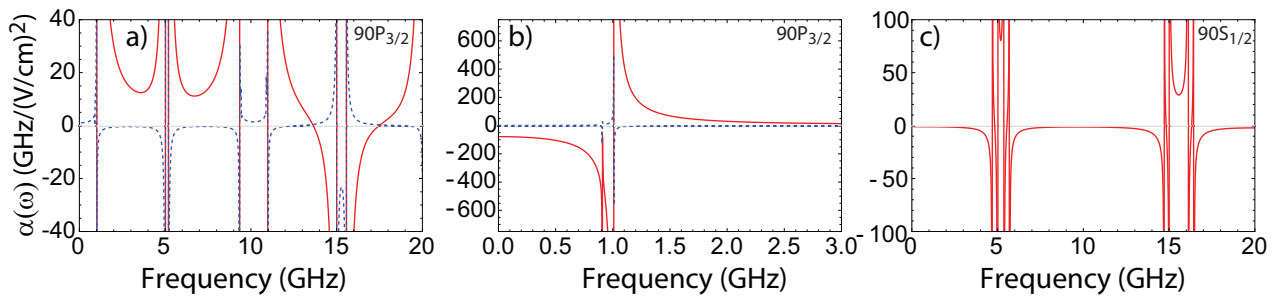}
    \caption{ac polarizability of the $90P_{3/2}, m_J = 3/2$ and $90S_{1/2}, m_J = 1/2$ Rydberg states. a) shows the  $90P_{3/2}$ state with the lower frequency range in b).  c) shows the  $90S_{1/2}$ state. The solid red curve is the scalar polarizability and the dashed blue curve is the tensor polarizability. }
    \label{fig:ACStark}
\end{figure*}

To obtain these quasi-energies, we first calculate the Floquet Hamiltonian. This Hamiltonian is formed by performing a Fourier expansion on the system's periodic Hamiltonian. The result is an infinite-dimensional matrix whose indices iterate over the atomic states and the Fourier components. The Floquet Hamiltonian is then derived from this matrix in the form \cite{TSHo1984}
\begin{eqnarray}
\braket{\alpha n_1 n_2 \left|H_{\rm F} \right| \beta m_1 m_2 } &=& \braket{\alpha n_1 n_2 \left|H \right| \beta m_1 m_2 }\nonumber\\
& +& (n_1 \omega_1 + n_2 \omega_2) \delta_{\alpha \beta} \delta_{n_1 m_1} \delta_{n_2 m_2}  ,\nonumber\\ \label{eq.Floquet}
\end{eqnarray}
where $\alpha,  \beta$ denote atomic states, $n_i, m_i$ are Fourier components,   $\ket{\alpha n_1n_2}$, $\ket{\beta m_1m_2}$ are the Floquet states,  and $\omega_i$ are the frequencies of the periodic part of the Hamiltonian. The eigenvalues of $H_F$ are the quasi-energies $q_\alpha$ of the modes of the periodic system. The matrix elements $\braket{\alpha n_1 n_2 \left|H \right| \beta m_1 m_2 }$ are 
\begin{eqnarray}
\braket{\alpha n_1 n_2 \left|H \right| \beta m_1 m_2 } &=& \delta_{n_1,m_1} \delta_{n_2,m_2} \braket{\alpha|H_0|\beta} \nonumber\\
&+& \sum_{i=1,2} \delta_{\left|n_i-m_i\right|,1} \; \braket{\alpha \left| z \right| \beta} F_{i}e^{\imath\phi_i} .\nonumber
\end{eqnarray}
We will assume that $\phi_1=\phi_2=0$. 
The Floquet method assumes that the Hamiltonian is periodic in time, which is the case for a single-tone field. In our case, however, we want to consider two-tone fields. These fields are not in general periodic, particularly if the ratio of the two fields is not a rational number. However, in \cite{TSHo1983}, it was shown that by treating the Floquet matrix for one of the fields as a time-independent infinite-dimensional Hamiltonian from which the second tone's Floquet matrix is calculated, it is possible to derive a form of the Floquet formalism that does not explicitly depend on the overall periodicity of the full system, and is thus applicable even if the system is not explicitly periodic in time.

 Because the Floquet Hamiltonian is infinite-dimensional, it must be truncated for the calculation to be possible. The dimension for truncation is selected by varying the number of included Fourier components and testing for convergence. For the parameters we are using, convergence occurs with three Fourier components of each frequency on each side of the zero component for a total of $7^2=49$ components. 
The Floquet Hamiltonian can then be diagonalized to obtain the quasi-energies. The process of calculating the Floquet matrices and diagonalizing them must be repeated for every dc electric field over the range being considered, so this calculation can be computationally intensive, but is easily parallelizable\cite{poldress2017}.

The ac fields that are applied as part of the polarizability cancelling scheme induce an ac Stark shift in the atoms. The ac polarizability of the $90P_{3/2},m_J = 3/2$ and $90S_{1/2}, m_{J=1/2}$ states as a function of frequency are shown in Fig. \ref{fig:ACStark}. While there are some zero crossings in the ac polarizability which would allow us to also remove sensitivity to any variations in the ac field amplitude, none of these frequencies are suitable for cancelling the $\state{90P}{3/2}$ and $\state{90S}{1/2}$ dc polarizabilities. Since the ac field amplitude can be well stabilized experimentally (as opposed to the surface dc fields), fluctuations in the ac field amplitude are not expected to be an experimental limitation.

Useful values of the frequency and amplitude of the dressing fields were found by solving  for the dressed energies $E_{90S_{1/2}}, E_{90P_{3/2}}$ and searching for parameters that minimized the energy variation with respect to the dc field amplitude $F_{\rm dc}$. Since the polarizability of the atomic ground state is negligible compared to the Rydberg polarizability, ground state shifts were not included in the calculation.

\section{Numerical results}
\label{sec.results}

We proceed to present numerical results for simultaneous stabilization of two states $\state{90P}{3/2}$ and $\state{90S}{1/2}$ with respect to dc field fluctuations (Sec. \ref{sec.2f2state})
and higher order stabilization of only the $\state{90P}{3/2}$ state (Sec. \ref{sec.2f1state}). The effect of ac dressing on Rydberg interactions is shown in Sec. \ref{sec.interactions}.

\subsection{Simultaneous stabilization of two states}
\label{sec.2f2state}

The results  in this section used $F_1=F_2= 26 ~\rm mV/cm$, $\omega_1 = 2\pi\times 800$ MHz, and $\omega_2=2\pi\times 2172$ MHz. The detunings with respect to the dominant transitions are $-144 $ MHz relative to the $\state{90P}{3/2} \leftrightarrow \state{89D}{5/2}$ transition for $\omega_1$ and \numunit{+8}{MHz} from the $\state{90S}{1/2} \leftrightarrow \state{88D}{5/2}$ two-photon transition for $\omega_2.$
 These frequencies are not at zero crossings of the ac polarizability, and as a result the atoms are sensitive to variations in the ac field. The ac polarizability for the $\state{90P}{3/2}$ state is \numunit{-218}{\GHzpVpcm} due to the \numunit{800}{MHz} field and \numunit{25.1}{\GHzpVpcm} due to the \numunit{2172}{MHz}, and the ac polarizability for the $\state{90S}{1/2}$ state is \numunit{-1.1}{\GHzpVpcm} due to the \numunit{800}{MHz} field, and \numunit{-1.4}{\GHzpVpcm} due to the \numunit{2172}{MHz} field. 

\begin{figure}[!b]
    \centering
    \includegraphics[width=0.48\textwidth]{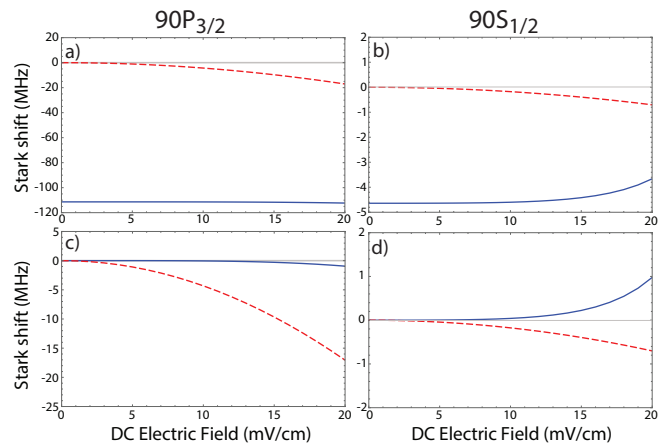}
    \caption{Energy shift of the $90P_{3/2}, m_J=3/2$ and $90S_{1/2}, m_J=1/2$ states due to a dc field with and without ac dressing.  (a,b) Undressed dc Stark shift  (dashed red curve) and shift with two-frequency microwave dressing (blue curve). The offset in the dressed state is due to the ac Stark shift from the microwave field. (c,d) The ac Stark shift due to the microwave field is subtracted.  }
    \label{fig:90PStark}
\end{figure}

 Figure \ref{fig:90PStark} a),c) shows the dc Stark shift of the $\state{90P}{3/2}$ levels in response to a dc field parallel to the microwave polarization. The zero of the vertical axis in Fig. \ref{fig:90PStark}a is the undressed $\state{90P}{3/2}$ energy. The energy with zero dc field is shifted by \numunit{-111}{MHz} due to the non-zero ac polarizability in the two-frequency field, with $\sim 99\%$ of the shift due to the \numunit{800}{MHz} component. Additionally, because the ac Stark shift depends on $m_J$, the $m_J$ components of the $\state{90P}{3/2}$ state are shifted by different amounts and are no longer degenerate at zero dc field. 

In these conditions the sensitivity to the dc electric field when the dc  field is parallel to the ac field polarization, is reduced by a factor of 1600 compared to the undressed state. In the case of the $\state{90P}{3/2}, m_j=3/2$ state, the scalar polarizability $\alpha_{0,\rm dc}$ is not reduced to zero, but instead is tuned by the ac field to be nearly equal and opposite in magnitude to the 
tensor polarizability $\alpha_{2,\rm dc}$. The changes in the polarizability are summarized in Table \ref{Tab:poltab}.

The dc Stark shift for the $\state{90S}{1/2}$ state is similarly shown in Fig. \ref{fig:90PStark} b),d). With the same ac field, a suppression of the second-order polarizability by a factor of 95 is possible (\numunit{-0.037}vs. \numunit{3.5}{\GHzpVpcm}). Due to the ac Stark shift, there is an offset of \numunit{-4.8}{MHz} at zero dc field relative to the undressed state. It is evident from the figure that the Stark shifts are not well described by a quadratic dependence on the dc field strength. This is because the calculation is based on diagonalization of the full Hamiltonian and therefore results in a hyperpolarizability that accounts for all orders of the electric dipole interaction. The specified reduction in the polarizability is found from fitting a quadratic function to the low field portion of the curves.  

\begin{table*}[!t]
\caption{Scalar and tensor dc polarizabilities with and without dressing fields.  The last line gives the values for the higher order cancellation of $90P_{3/2}$ described in Sec. \ref{sec.2f1state}. The last column is the ac shift due to the dressing fields. }
\label{Tab:poltab}
\begin{tabular}{|l|c|c|c|c|c|}
\hline
State     &  $F_1$ & $F_2$ &   $\alpha_{0,\rm dc}$   & $\alpha_{2,\rm dc}$  &   ac Stark shift \\ 
     &    &  &   ($\GHzpVpcm$)  & ($\GHzpVpcm$)   &   (MHz) \\ \hline

$90S_{1/2},m_J=1/2$ & - & - & 3.50 & 0 & 0 \\ \hline

$90S_{1/2},m_J=1/2$ & 26 mV/cm, 800 MHz & 26 mV/cm, 2172 MHz & -0.036 & 0 & -4 \\ \hline

$90P_{3/2},m_J=3/2$ & - & -   &  95.5   &   -8.27   & 0 \\ \hline

$90P_{3/2},m_J=3/2$ & 26 mV/cm, 800 MHz & 26 mV/cm, 2172 MHz &  1.29   &   -1.08   & -111   \\ \hline

$90P_{3/2},m_J=3/2$ & 28 mV/cm, 720 MHz & 10 mV/cm, 5600 MHz &  0.059   &   0.137   & -381 \\   \hline
\end{tabular}

\end{table*}

In a zero bias field, fluctuations in the dc field cannot be assumed to be parallel to the microwave polarization. Thus the angular dependence of the polarizability must be considered as well. This is especially true for the $90P_{3/2}, m_J=3/2$ state, which possesses a non-zero tensor polarizability, introducing an angular dependence of the Stark shift in the form
\begin{equation}
\Delta U = -\frac{F_{\rm dc}^2}{2} \alpha_{0,\rm dc} + \frac{F_{\rm dc}^2(1-3\mathrm{cos}^2\theta)}{4} \frac{3m_J^2-J(J+1)}{J(2J-1)}\alpha_{2\rm ,dc} ,\nonumber
\end{equation}
where $\theta$ is the angle between the quantization axis (set to be along the direction of the microwave polarization) and the dc electric field axis.

\begin{figure}[!t]
    \centering
    \includegraphics[width=0.48\textwidth]{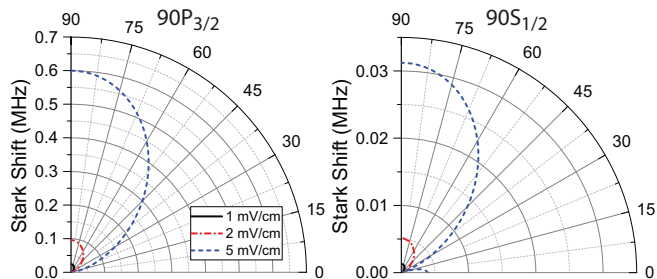}
    \caption{Polar plots of the energy shift of the $90P_{3/2}, m_J=3/2$ and $90S_{1/2}, m_J=1/2$ states dressed with a two-frequency microwave field with  dc electric field strength of $1, 2, 5$ mV/cm  and variable direction. 0 deg. corresponds to parallel dc field and ac field polarization.}
    \label{fig:PolarPlot90P}
\end{figure}

\begin{figure}[!t]
    \centering
    \includegraphics[width=0.48\textwidth]{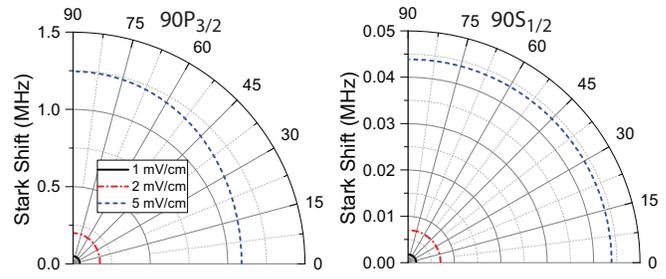}
    \caption{Polar plot of the energy shift of the $90P_{3/2},m_J=3/2$ and $90S_{1/2},m_J=1/2$ states without microwave fields and with varying dc electric field direction. The three traces correspond to dc field amplitudes of $1, 2, 5$ mV/cm. Because the quantization axis is held fixed and the dc field direction allowed to vary, there is a slight anisotropy induced in the $90P_{3/2}, m_J=3/2$ state by the tensor polarizability. 0 deg. corresponds to parallel dc field and ac field polarization.}
    \label{fig:PolarPlot90PnoRF}
\end{figure}

The angular dependence of the Stark shift of the $90P_{3/2}, m_J=3/2$ and $90S_{1/2}, m_J=1/2$ states in the two-frequency  microwave field is shown in Fig. \ref{fig:PolarPlot90P}.  For comparison, the angular dependence of the  state energies with no microwave dressing field is shown in Fig. \ref{fig:PolarPlot90PnoRF}. The polarizability cancellation varies widely with angle, and is least effective for perpendicular ac and dc fields. Nevertheless there is suppression  at all angles. For the $90P_{3/2}, m_J=3/2$ state the suppression has been optimized at 0 deg. dc field direction. While this gives  suppression of the $90S_{1/2}, m_J=1/2$ polarizability at 0 deg. by a factor of 95, the suppression is even stronger at approximately 20 deg. field direction. Although we have not done it, the dependence on direction for small dc fields could in principle be further reduced by choosing dressing parameters that minimize the polarizability at finite dc bias field. Small fluctuating fields would then have a minimal effect on the field direction. 

\subsection{Cancellation of the fourth order polarizability}
\label{sec.2f1state}

\begin{figure}[!t]
    \centering
    \includegraphics[width=0.4\textwidth]{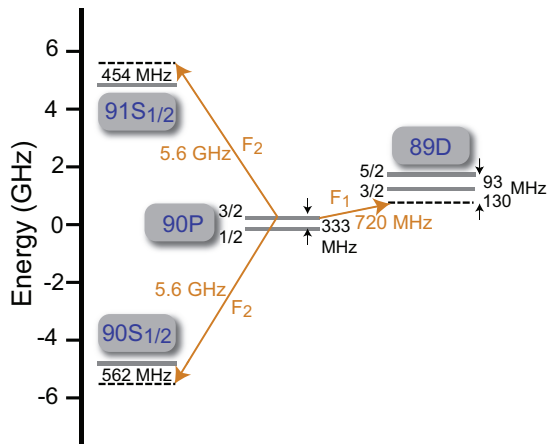}
    \caption{Level diagram and dressing fields for single-state higher order polarizability nulling. The energy scale is relative to the $\state{90P}{3/2}$
level. Microwave fields with frequencies \numunit{720}{MHz} and \numunit{5600}{MHz}, depicted in the diagram as dotted arrows, couple the $\state{90P}{3/2}$ state off-resonantly to the $\state{89D}{5/2}$ state and the $\state{91S}{1/2}$ state. Some coupling also occurs to the $\state{90S}{1/2}$ state, but the detuning is larger by \numunit{110}{MHz}.}
    \label{fig:leveldiagramP}
\end{figure}

The polarizability-reduced states described above have a minimized second-order dependence on the dc electric field, but as can be seen particularly in the $\state{90S}{1/2}$ state, the state mixing that causes this cancellation can also introduce a larger fourth-order dependence which dominates at dc fields larger than 15 mV/cm. Thus it would be useful to be able to cancel both the second- and fourth-order terms for a single state simultaneously. To do this for the $\state{90P}{3/2}$ state, we used a two-frequency microwave field with a component of frequency \numunit{720}{MHz} and amplitude 28 mV/cm and a component of frequency \numunit{5.60}{GHz} and amplitude 10 mV/cm. The first component couples the $\state{90P}{3/2}$ state to $\state{89D}{5/2}$, and the second component couples the $\state{90P}{3/2}$ state to $\state{90S}{1/2}$ and $\state{91S}{1/2}$. The level diagram and coupling fields are shown in Fig. \ref{fig:leveldiagramP}.

\begin{figure}[!t]
    \centering
    \includegraphics[width=0.4\textwidth]{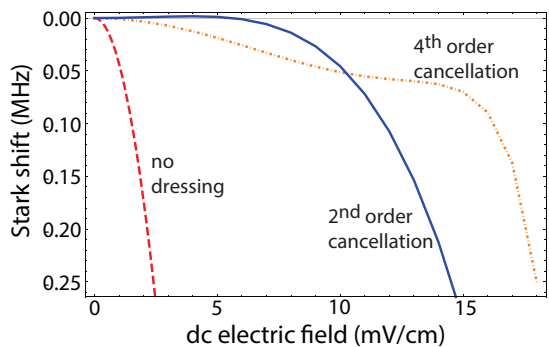}
    \caption{The dc Stark shift of the $90P_{3/2}, m_J=3/2$ state dressed by a two-frequency field with a component of frequency \numunit{720}{MHz} and amplitude \numunit{28}{\mVpcm} and a component of frequency \numunit{5.60}{GHz} and amplitude 10 mV/cm (orange dots). The solid blue line is the same state dressed with the two-frequency field from Sec. \ref{sec.2f2state}. The dashed red line is the Stark shift with no dressing fields.}
    \label{fig:TwoFreq}
\end{figure}

The Stark shift of the dressed $\state{90P}{3/2}$ state in this two-frequency field is shown in Fig. \ref{fig:TwoFreq}. The offset due to the ac Stark shift is subtracted from the dressed curves. At fields less than 10 mV/cm, the reduction in the dc Stark shift is somewhat worse than in the dual-state cancellation scheme of the previous section. However, when the dc field is between 11 mV/cm and 17 mV/cm, the reduction in the dc Stark shift is greater. At
17 mV/cm there is an avoided crossing with another Stark curve which increases the dc Stark shift at higher fields.

\begin{figure}[!t]
    \centering
    \includegraphics[width=0.4\textwidth]{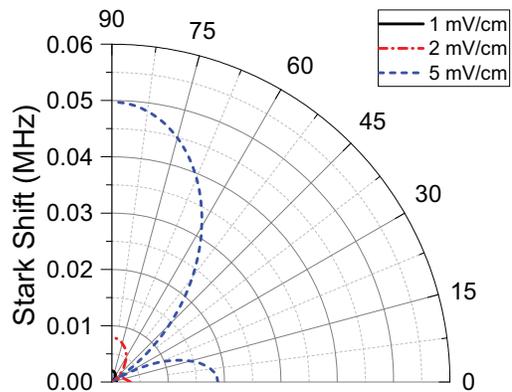}
    \caption{Polar plot of the $90P_{3/2}, m_J=3/2$ state dressed with a \numunit{28}{\mVpcm} 720 MHz and \numunit{10}{\mVpcm} 2172 MHz two-frequency microwave field with varying dc electric field. The three traces correspond to dc field amplitudes of \numunit{1}{\mVpcm}, \numunit{2}{\mVpcm}, and \numunit{5}{\mVpcm}.}
    \label{fig:PolarPlot90P2freq}
\end{figure}

As in the other two-frequency dressed state case, the dc Stark shift depends on the angle between the dc field and the ac field. This variation is shown in Fig. \ref{fig:PolarPlot90P2freq}. Though there is angular dependence, the sensitivity to the dc field is reduced by a factor of at least 25 at all angles compared to the undressed case, shown in Fig. \ref{fig:PolarPlot90PnoRF}.  Although the performance at 0 deg. is worse than that of the Sec. \ref{sec.2f2state} dressing shown in Fig. \ref{fig:PolarPlot90P}, the performance at 90 deg. is an order of magnitude better. 

\subsection{Interaction of dressed Rydberg states}
\label{sec.interactions}

It is also important to consider the effect of the dressing fields  on Rydberg-Rydberg interactions. To check this we calculated the interaction energy for a pair of atoms in the $90P_{3/2},m_J=3/2$ state  in the absence of microwave  fields and in the presence of the two-frequency field from Fig. \ref{fig:TwoFreq} using both the Floquet method and  perturbation theory. The Rydberg-Rydberg interactions were calculated using the method described in \cite{Walker2008}, including dipole-dipole terms and neglecting higher-order terms which are not significant at long range.

The two-atom dipolar coupling is calculated as  $V_{\rm dd}=C_3/R^3$ where $R$ is the atomic separation and the $C_3$ coefficient is given by the general expression
$C_3=\bra{\alpha_1}\bra{\alpha_2}\hat C_3 \ket{\alpha_1}\ket{\alpha_2}$ with 
\begin{eqnarray}
\hat C_3&=&-\frac{e^2}{4\pi\epsilon_0}\frac{\sqrt6 (4\pi)^{3/2}}{3\sqrt5 } \sum_{M,q}
(-1)^MC_{1,q,1,-M-q}^{2,-M}\nonumber\\
&\times& Y_{2,M}({\bf n})(rY_{1,q})^{(1)}(rY_{1,-M-q})^{(2)}.\nonumber\end{eqnarray}
Here $e$ is the electronic charge, $\epsilon_0$ is the permittivity of free space,
and $\bf n$ is a unit vector along the molecular axis. 
The superscripts denote the coordinates of atom 1 and atom 2.
For the undressed basis and the perturbation basis $\ket{\alpha}$ is defined by the quantum numbers $\{n,l,j,m_j\}$. For the Floquet basis the
quantum numbers are $\{n,l,j,m_j,n_1,n_2\}$, where $n_i$ is the Floquet order of the  the $i^{\rm th}$ field component of the state, as defined in Eq. (\ref{eq.Floquet}).

\begin{figure}[!t]
    \centering
    \includegraphics[width=0.48\textwidth]{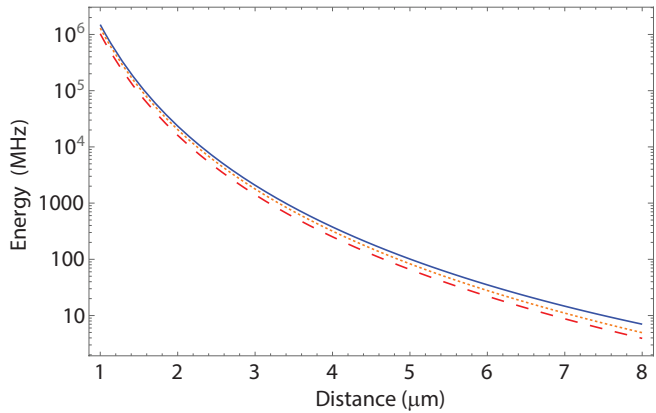}
    \caption{Comparison of the undressed Rydberg-Rydberg potential for a $90P_{3/2}, m_J=3/2+90P_{3/2}, m_J=3/2$ pair (blue, solid curve), the Rydberg-Rydberg interaction with a two-frequency field calculated using the Floquet method (red, dashed), and the Rydberg-Rydberg interaction with a two-frequency field calculated using perturbation theory (orange, dotted).}
    \label{fig:LongRangeRydRyd}
\end{figure}

The interactions were calculated for $90P_{3/2}, m_J=3/2 +90P_{3/2}, m_J=3/2 $ with a \numunit{10}{\mVpcm} dc field in three cases: 1) in the absence of an RF field, 2) with two-frequency dressing fields at 720 MHz and 5.6 GHz calculated in the Floquet basis, and 3) with the same dressing fields  calculated using perturbation. These interactions were then fit to determine the $C_6$ coefficient in each case. In the first case, the $C_6$ coefficient for this state was \numunit{-1.49}{THz\;\mu m^6}. In the Floquet case, it was \numunit{-1.03}{THz\;\mu m^6}, and in the perturbation case, it was \numunit{-1.30}{THz\;\mu m^6}. This shows that the RF field has a measurable, but not problematic, effect on the long-range interactions of the Rydberg pair as can be seen 
 in Fig. \ref{fig:LongRangeRydRyd}. 

\section{Conclusions}
\label{sec.summary}

In summary we have shown that using  two-frequency microwave dressing the dc polarizability of the $\state{90S}{1/2}, m_J=1/2$ and $\state{90P}{3/2}, m_J=3/2$ states in Cs can be suppressed by factors  of 95 and 1600 respectively for dc fields parallel to the quantization axis and the polarization of the ac dressing field. This extends previous work on reduction of the differential shift of neighboring Rydberg states\cite{Hyafil2004,Mozley2005,Jones2013,Ni2015}. We anticipate that the ability to greatly reduce the polarizability will be important for quantum gate experiments with  Rydberg atoms where stability of the ground-Rydberg energy separation is a requirement for achieving high gate fidelity.  

The polarizability suppression was optimized for a dc field parallel to the quantization axis. For fields in other directions the effectiveness of the suppression is reduced, as shown in Fig. \ref{fig:PolarPlot90P}. It is possible to further reduce the polarizability at relatively large dc fields ($\sim 10-17~\rm  mV/cm$) by applying a two-frequency ac field to cancel both the second- and fourth-order polarizabilities of the $\state{90P}{3/2}$ state. 
This  has the additional benefit of providing strong suppression independent of the direction of the dc field. 
There is some effect of the dressing field on Rydberg-Rydberg interactions, but it is small enough that this method may be useful in cases where the pair interaction is important.

While our calculations were performed for specific Cs atom Rydberg states we expect that the method can be readily adapted to other states and other atomic species. 
It is also possible that improved suppression factors can be achieved by adding more dressing frequencies, beyond the case of two frequencies considered here. 
A limitation of this method is that it requires relatively small ($< 15 ~\rm mV/cm$) background dc electric fields, which means that for near-surface experiments, some effort will be required to control the background electric field. Methods such as those in \cite{Hermann-Avigliano2014} and \cite{Sedlacek2016} could be combined with the method presented in this work to further reduce sensitivity to stray electric fields.

\begin{acknowledgments}
This research was supported by Army
Research Office Contract No. W911NF-16-1-0133 and by
the US Army Research Laboratory Center for Distributed
Quantum Information through Cooperative Agreement No.
W911NF-15-2-0061. 

This research was performed using the compute resources and assistance of the UW-Madison Center For High Throughput Computing (CHTC) in the Department of Computer Sciences. The CHTC is supported by UW-Madison, the Advanced Computing Initiative, the Wisconsin Alumni Research Foundation, the Wisconsin Institutes for Discovery, and the National Science Foundation, and is an active member of the Open Science Grid, which is supported by the National Science Foundation and the U.S. Department of Energy's Office of Science. 

\end{acknowledgments}


\input{floquet_dressing_resub.bbl}

\end{document}

%% file: floquet_dressing_resub.bbl
%

%% file: floquet_dressing_resub.bbl
\begin{thebibliography}{43}
\bibitem [{pol()}]{poldress2017}%
  \BibitemOpen
  \href@noop {} {}\bibinfo {note} {Calculations for $m_J=1/2$ with the field
  along the quantization axis used a basis size of 1170 thus requiring
  diagonalization of $1170\times 1170$ matrices. The large basis size arises
  from including seven Fourier components and all states differing from the
  initial state by $\delta n^*\le 3.5$ and with $L\le 5$. For cases with the
  field not along the quantization axis there is coupling to other $m_J$ values
  which required a basis size of 5851 states. Computation time on the HTCondor
  cluster at University of Wisconsin-Madison was 3(24) hours for a single case
  of field parallel(not parallel) with the quantization axis. COmputations with
  100 parameter values were executed in parallel using 100 nodes of the
  cluster, with each node having 20 computational cores. For the
  Rydberg-Rydberg interaction calculation of Fig. 9 the matrix size depended on
  where the cutoff for the energy spacing was, and typically required $\sim
  5000$ states. Run times were 12-24 hours for a single set of
  parameters.}\BibitemShut {Stop}%
\end{thebibliography}
